\documentclass[12pt]{iopart}

\usepackage{makeidx}
\usepackage{amsfonts}
\usepackage{amssymb}
\usepackage{enumerate}
\usepackage{latexsym}
\usepackage{bbm}
\usepackage{graphicx}
\usepackage{graphics}
\usepackage{float}

\newcommand{\beq}{\begin{equation}}
\newcommand{\eeq}{\end{equation}}
\newcommand{\beqa}{\begin{eqnarray}}
\newcommand{\eeqa}{\end{eqnarray}}

\newcommand{\ket}[1]{\left\vert #1 \right\rangle}

\newcommand{\vect}[1]{\bi{#1}}
\newcommand{\env}{\psi}
\newcommand{\drei}{}
\newcommand{\zwei}{\prime}
\newcommand{\eins}{{\prime\prime}}
\newcommand{\einsB}{\prime}
\newcommand{\nichts}{{\prime\prime\prime}}

\begin{document}

\title{On the efficiency of quantum lithography}

\author{Christian Kothe$^{1,2}$, Gunnar Bj\"{o}rk$^1$, Shuichiro Inoue$^3$ and Mohamed Bourennane$^2$}
\address{$^1$ School of Information and Communication Technology, Royal Institute of Technology (KTH), Electrum 229, SE-164 40 Kista, Sweden}
\address{$^2$ Department of Physics, Stockholm University, SE-109 61 Stockholm, Sweden}
\address{$^3$ Institute of Quantum Science, Nihon University, 1-8-14 Kanda-Surugadai, Chiyoda-ku, Tokyo 101-8308, Japan}
\ead{kothe@kth.se}

\date{\today}

\begin{abstract}

Quantum lithography promises, in principle, unlimited feature resolution, independent of wavelength. However, in the literature at least two different theoretical descriptions of quantum lithography exist. They differ in to which extent they predict that the photons retain spatial correlation from generation to the absorption, and while both predict the same feature size, they differ vastly in predicting how efficiently a quantum lithographic pattern can be exposed.

Until recently, essentially all experiments reported have been performed in such a way that it is difficult to distinguish between the two theoretical explanations. However, last year an experiment was performed which gives different outcomes for the two theories. We comment on the experiment and show that the model that fits the data unfortunately indicates that the trade-off between resolution and efficiency in quantum lithography is very unfavourable.
\end{abstract}


\maketitle

\section{Introduction}

For a long time it was believed that the resolution of optical systems was limited by the so called Rayleigh criterion \cite{BW}, which roughly states that the resolution of the system is diffraction limited by the half of the wavelength $\lambda$ of the used light. Hence, it was believed that only by going to shorter and shorter wavelengths one could improve the resolution substantially. However, going from the visibile to UV-light, x-rays, or gamma radiation means using light with such a high photon energy that the radiation may damage or destroy the object one wants to expose or investigate. Furthermore, it is difficult to build optical systems, since ordinary lenses are no longer transparent in these wavelengths regions, and even gases start to absorb the light quite substantially. Moreover, building mirrors are also challenging due to absorption and very small refractive index differences between different materials. This makes it complicated and expensive to build Rayleigh limited optical systems for wavelengths shorter than about 200 nm.

There have been proposals for ``classical'' systems which achieve higher resolution than the Rayleigh limit \cite{YV, BB1, BB2}, but such lithographic schemes suffer from visibility problems and they cannot simultaneously generate a sub-Rayleigh feature size and a sub-Rayleigh feature separation, at least not until new kinds of detectors are developed \cite{OKS,HMS, SHZ}.

It therefore came as a relief when Boto \etal showed that entanglement could be used to beat the diffraction limit given by the Rayleigh criterion \cite{Boto}. They showed that the use of $N$ entangled photons would allow one to surpass the diffraction limit by a factor of $1/N$ compared to the Rayleigh criterion. Lithographic systems would potentially profit substantially from the idea of Boto \etal by allowing the generation of features arbitrarily smaller than the Rayleigh limit \cite{DCS,BSS1,BSS2,DSS,Shih,FFO,PDV,GLM}. To make these systems work in practice, there is still a plethora of problems to solve. For example, one has to find a material which changes its behaviour only when absorbing $N$ photons \cite{YV, CSO, ACB}, and one needs also to optimise the absorption rate of this material \cite{PWA}. Unfortunately, there are limits to the extent this is feasible \cite{Tsang}. Additional difficulties involve the rather peculiar, multi-photon, entangled photon states that need to be generated \cite{SVM, GCD}. Alternatively,  other ways to generate indistinguishable $N$-tuples of photons have to be developed \cite{TBZ}. There have been investigations for the case of multiple frequencies or in the broadband limit \cite{KY,CW}. For some schemes even photon losses may become an issue \cite{GAW}. Furthermore one has to think about how to generate arbitrary patterns \cite{BSS1,BSS2,DSS,KBA}. An open question that we address in this paper is if systems built on this principle also would work \textit{efficiently} in practice. The main aspect is whether or not the time needed to irradiate a lithographic film would be short enough to make this method efficient.

We start our paper in section \ref{sec:theo} by describing two different models for most of the experiments hitherto reported in the field of quantum lithography, but where one of the explanations predicts a favourable time scaling whereas the other one does not. We then discuss an experiment in section \ref{sec:exp} which would give different outcomes for the two explanations and which thereby could decide which of the explanations is correct (or at least applicable to this kind of experiment) and how the time for imaging or radiation scales with the number of photons $N$ per state and the number of image ``pixels'' $S$. The outcome of an experiment recently performed by Peeters \etal \cite{PRE} allows us then to deduce which of the explanations seems to be applicable, and the result suggests that the time-scaling behaviour of such a quantum lithography scheme is, unfortunately, very unfavourable. In section \ref{sec:rel} we relate our paper to other papers in the field before we conclude our results in section \ref{sec:concl}. We have also two appendices, one where we give a thorough mathematical analysis and derivation of our model and one where we look in more detail into the original proposal of Boto \etal.

\section{Different theoretical explanations}
\label{sec:theo}

\begin{figure}
\center
\includegraphics[angle=0, scale=.7]{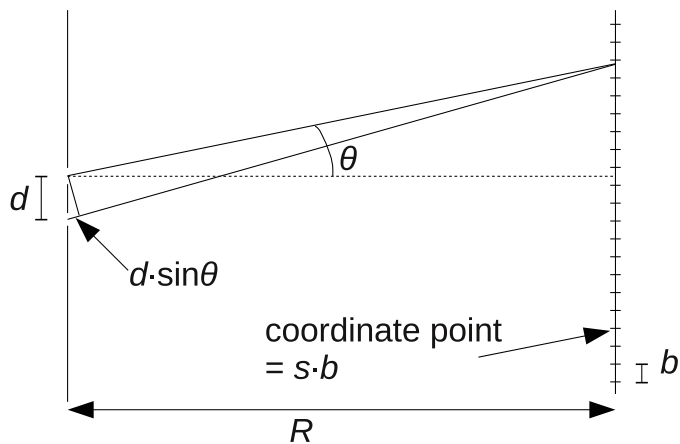}
\caption{Setup for quantum imaging or lithography for $N=2$. Each of the in total $S+1$ coordinate points on the screen to the right consists of a detector with width $b$ which detects two-photon absorption. The double slit on the left has a distance $d$ between the slits and is at a distance $R$ to the screen.}
\label{Fig:experiment1}
\end{figure}

Imagine a setup as depicted in figure \ref{Fig:experiment1}. A light beam of wavelength $\lambda$ is incident on a double slit with slit distance $d \gg \lambda$. Due to diffraction at the slit, the light will produce an interference pattern on a screen at a distance $R\gg d$ behind the slit. Assume that our screen now consists of $S+1$ detectors, each of the same width $b$. Below, we have assumed that $S$ is even (representing an odd number of detectors). However, the conclusions we draw are equally valid if an even number of detectors are used. The detectors cover the whole length of the screen and $b$ is small enough to clearly be able to resolve the interference pattern, i.e., the distance between two intensity peaks in the detector plane is many times larger than $b$. A photon impinging on the double slit can either pass through the upper or through the lower slit, i.e., the state at the slit will be
\beq
\label{eq:1}
\ket{\Psi_{{\rm 1,sl}}}=\frac{1}{\sqrt{2}}\left(\ket{1}_{upper} \otimes \ket{0}_{lower}+\ket{0}_{upper} \otimes\ket{1}_{lower}\right).
\eeq
Since we have decided to remain ignorant as to which path the photon took through the double slit, but only detect the photon with the detectors at the screen, we will there have the state
\beqa
\nonumber
\ket{\Psi_{{\rm 1,sc}}} & = & \frac{1}{{\cal N}_1}\sum\limits_{s=-\frac{S}{2}}^{\frac{S}{2}}\left\{ \rme^{\rmi k r\left(s\right)} \hat{a}_s^\dagger + \rme^{\rmi k \left[r\left(s\right)+\Delta r\left(s\right)\right]} \hat{a}_s^\dagger \right\}\ket{0}\\
\label{eq:2} & = &
\frac{1}{{\cal N}_1}\sum\limits_{s=-\frac{S}{2}}^{\frac{S}{2}}\rme^{\rmi k r\left(s\right)}\left[1+\rme^{\rmi k \Delta r\left(s\right)}\right]\hat{a}_s^\dagger\ket{0}, \eeqa
where ${\cal N}_1 = \sqrt{2(S+1)}$, the summation is over all detector modes $s$, $k$ is the wave vector of the incident light, $r\left(s\right)=\sqrt{R^2+s^2 b^2}\approx R\left(1+\frac{s^2 b^2}{2R^2}\right)$ and $\Delta r\left(s\right)=d\sin \theta \approx d\theta=dbs/R$. (For the normalisation in expression (\ref{eq:2}) we have for simplicity assumed that the pattern falling on the $S+1$ detectors contains exactly an integer number of fringes.) The ensuing measurement probability at the detection plane is
\beq
P_{{\rm 1,sc}}(s) \propto \cos^2(\frac{k d b s}{2 R}).
\label{Eq: single photon}
\eeq
The distance between two peaks in the measured interference pattern is given by $\lambda_0 R/d$, where $\lambda_0$ is the used wavelength. In the equation above we have neglected the envelope function induced by the exact shape of each slit. Instead we have assumed that the photons have equal probability amplitude to hit any detector. This is a simplification justified by the fact that it does not alter our main results or conclusion.

If we look at the quantum lithography scheme the setup is slightly changed. We are not any longer interested in single photons but in ``bunches'' of $N$ photons. For most of our paper we restrict ourselves for practical reasons to $N=2$ photons, but our whole argumentations will be valid for any number $N$. To show that one can get a higher resolution (i.e., a shorter distance between peaks) in this setup, one wants to assure that two indistinguishable photons
pass through the same slit, i.e., that one has a two-photon, NOON state at the slits \cite{Boto}
\beqa
\label{eq:3}
\ket{\Psi_{{\rm 2,sl}}}=\frac{1}{\sqrt{2}}\left(\ket{2}_{upper} \otimes\ket{0}_{lower}+\ket{0}_{upper}\otimes\ket{2}_{lower}\right).
\eeqa
This can be implemented, e.g., by placing a short type-I crystal in front of the double slit and pumping this crystal so that it produces pairs of degenerate photons which leave the crystal collinearly. One also has to assure that the pump beam is broader than the distance between the slits and that the two slits are centered on the setup's optical axis. Note that the state at the screen is isomorphic to a Bell state, that is it has maximal transverse entanglement of its photon number. Also note that, since the slits can be viewed as the source illuminating the screen, an analysis of how this state is produced is not needed. Indeed, in  \ref{app:A} it turns out that even long crystals, narrow beams and a long distance between crystal and slits can produce $\ket{\Psi_{{\rm 2,sl}}}$. It suffices to conclude that the state $\ket{\Psi_{{\rm 2,sl}}}$ is the optimal 2-photon state for a demonstration of 2-photon interference fringes, giving unity two-photon interference visibility and interference fringe doubling as compared to the classical Young's experiment at the same wavelength of light.

The interesting question is what happens with the state $\ket{\Psi_{{\rm 2,sl}}}$ after that the photons have passed the double slit. One could use the concept of photonic de Broglie-waves \cite{JBC} to explain the effect of quantum imaging or lithography. Boto \etal \cite{Boto} analysed the probability for the two-photon state to arrive at a specific point on the screen depending on the intensity $I$ of the incoming light as follows: ``For two-photon absorption with entangled photon pairs, the absorption cross section scales as $I$... If the optical system is aligned properly, the probability of the first photon arriving in a small absorptive volume of space time is proportional to $I$. However, the remaining $N-1$ photons are constrained to arrive at the same place at the same time, and so each of their arrival probabilities is a constant, independent of $I$.''

Transferring this idea to the experiment in figure \ref{Fig:experiment1} would give the state
\beq
\label{eq:4}
\ket{\Psi_{{\rm B,sc}}}= \frac{1}{{\cal N}_{\rm B}} \sum_{s=-\frac{S}{2}}^{\frac{S}{2}}\rme^{\rmi 2k r(s)}\left[1 +\rme^{\rmi 2k \Delta r(s)}\right](\hat{a}_s^\dagger)^2\ket{0},
\eeq
where ${\cal N}_{\rm B}=\sqrt{4(S+1)}$. At the screen, the distance between two peaks $\lambda_0 R/(2 d)$ is only half as long as in (\ref{Eq: single photon}) and one could treat the two photons as a photonic de Broglie-wave. The implication of this state is that the two photons can hit essentially any detector (except those at the interference anti-nodes) but that both photons will hit the same detector.

Steuernagel \cite{Steuernagel}, however, opposed that description by stating that ``... it is not true that the first arriving photon greatly constrains the arrival location of the following ones ... Very few photons will be absorbed in one point since they typically arrive far apart.'' In a mathematical description his argumentation can be given the following description:
\beqa
\label{eq:5}
\ket{\Psi_{{\rm St,sc}}} & = & \frac{1}{{\cal N}_{\rm St}}\sum\limits_{s=-\frac{S}{2}}^{\frac{S}{2}}\sum\limits_{t=-\frac{S}{2}}^{\frac{S}{2}}  \left\{\rme^{\rmi k \left[r_1\left(s\right) + r_1\left(t\right)\right]} \hat{a}_s^\dagger\hat{a}_t^\dagger + \rme^{\rmi k \left[r_2\left(s\right)+r_2\left(t\right)\right]} \hat{a}_s^\dagger\hat{a}_t^\dagger \right\}\ket{0},
\eeqa
where
\beq
r_{1,2}\left(s\right)=R\left(1+\frac{s^2 b^2}{2R^2}\right)\mp\frac{db}{2R}s
\eeq
is the distance from the upper and lower slit, respectively, and ${\cal N}_{\rm St}=\sqrt{2}(S+1)$. Although one might argue that our derivation of (\ref{eq:4}) and (\ref{eq:5}) is somewhat short-handed or hand waving, a complete quantum mechanical analysis of the state from the pump beam to the screen leads to the same result. Such an analysis is given in \ref{app:A}.

Also Tsang opposed to the assertions of Boto \etal. He based his conclusion on a rather involved analysis, and found that the multi-photon absorption-rate should actually be lower for spatially entangled states, such as NOON-states (of which the state $\ket{\Psi_{{\rm 2,sl}}}$ is a $N=2$ realisation), than for classical states \cite{Tsang1}. However, in the following we will follow Steuernagel's somewhat simpler model.

The difference between the descriptions is that the photons retain a spatial correlation (``stick together'') in the description of Boto \etal, and hence always will arrive at the same position. In Steuernagel's model the photons will propagate independently of each other after the slit, and can hence arrive at different detectors. When looking at the case where both photons arrive at the same detector (that is, $s=t$, see below) the models coincide when it comes to the spatial behaviour and thus, by only looking at two-photon, same position events, one will se identical interference patterns, both compressed by a factor of $1/2$ as compared to the single photon interference, described in (\ref{eq:2}) and (\ref{Eq: single photon}).

Although predicting the same two-photon, same-postition absorption pattern, the two descriptions in (\ref{eq:4}) and (\ref{eq:5}) will give a quite different scaling behaviour for how efficient a two-photon exposure would be. In (\ref{eq:4}) both photons in every two-photon state passing through the slits arrive at the same detector, and going from 2 to $N$ photons in (\ref{eq:3}), the exposure efficiency will be invariant. (However, the absorption efficiency of any detector or material will in general be considerably lower, but this effect will be the same in both cases). Changing the number of detectors $S$ will only linearly increase the time one has to expose to get a certain number of $N$-photon events per detector. Hence, the ``exposure time'' of a quantum lithographic image will scale as $\tau_{\rm {B}}(S,N) \propto S$.

Equation (\ref{eq:5}), however, assumes that the arrival of the first photon at a certain detector does not in any way determine where the other photon will arrive and therefore, in the two-photon case, the exposure time $\tau_{{\rm St}}(S,2) \propto S^2$. The photons in an $N$ photon state will also hit the detectors in a random fashion, but only if all photons arrive at the same detector will the detector fire. Hence, the exposure time scaling for an $N$-photon state will be $\tau_{{\rm St}}(S,N) \propto S^N$. Hence, as pointed out by Steuernagel \cite{Steuernagel}, the exposure time will increase exponentially, and for realistic values of the number of detectors (pixels) and the efficiency of $N$-photon absorbers, this will essentially rule out
the practical use of quantum lithography. E.g., for a 5 $\times$ 5 pixel image the exposure time will multiply with 25 times going from a 2-photon state to a 3-photon state. For a more realistic image resolution of 100 $\times$ 100 pixels, the exposure time will increase by $10^4$ going from an $N$-photon state to an $N+1$ photon state. Therefore finding out which description is applicable is an important task to determine to what extent quantum lithography will be efficient.

\section{A conclusive experiment}
\label{sec:exp}

\subsection{Description of the experiment}
\begin{figure}
\center
\includegraphics[angle=0, scale=.7]{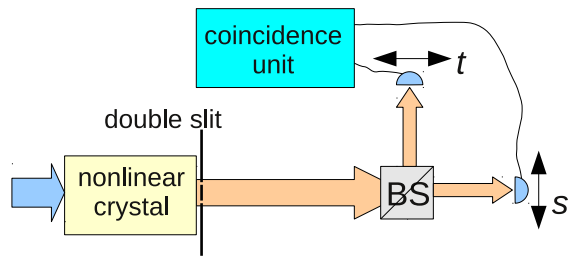}
\caption{Schematic experiment for measuring spatial correlations.
See the text for more details.} \label{Fig:experiment2}
\end{figure}

\begin{figure}
\center
\begin{tabular}{ c c }
\includegraphics[angle=0, scale=.5]{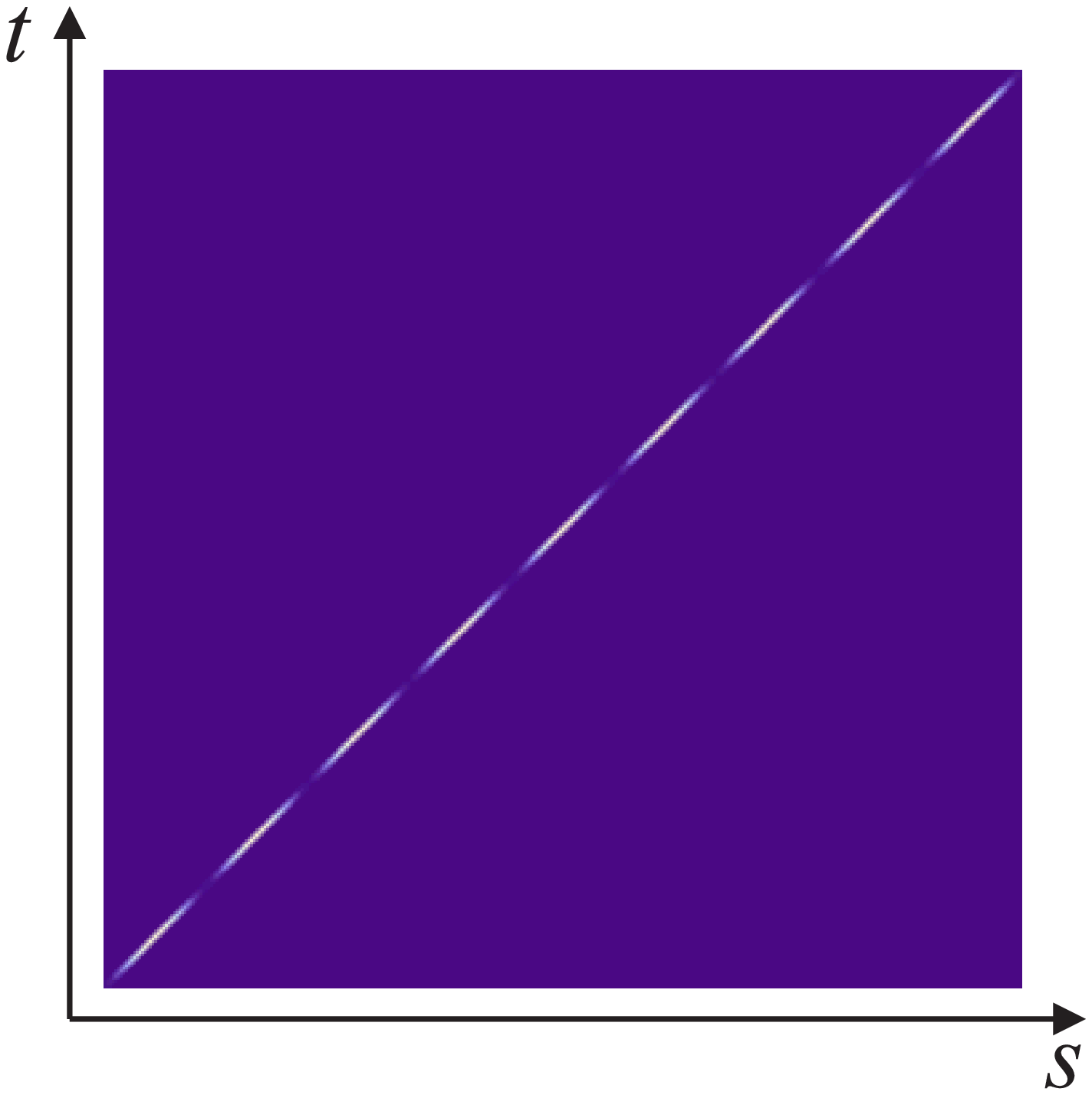} &
\includegraphics[angle=0, scale=.5]{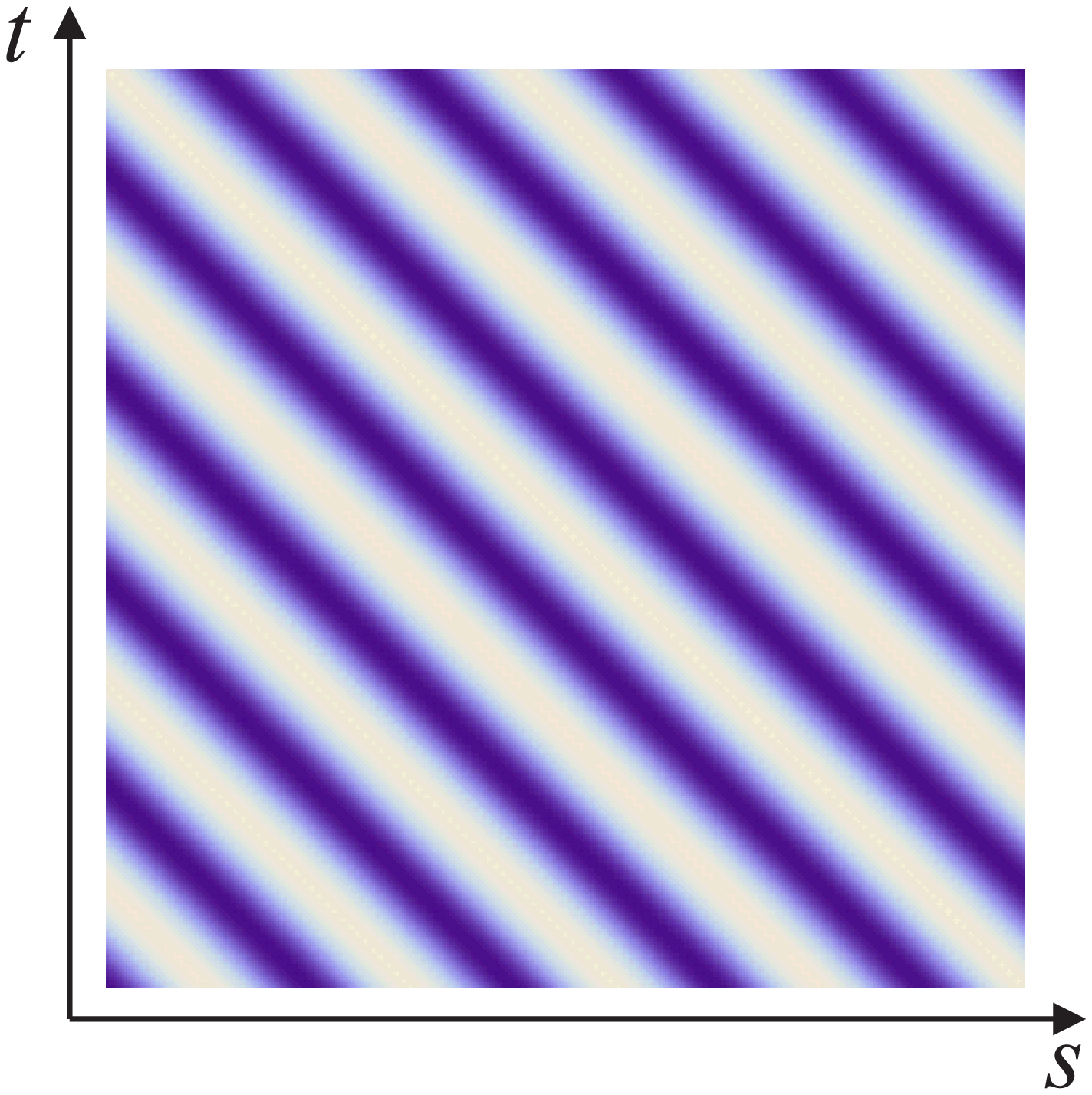}\\
(a) & (b) \\
\end{tabular}
\caption{Difference between the two theoretical explanations. In (a) the model of Boto \etal is displayed. In (b) the model of Steuernagel is illustrated. The axes correspond to the positions $s$ and $t$ of the two detectors in figure \ref{Fig:experiment2} respectively. The distance between two peaks along the line $s=t$ is half as long as it would be for the state in (\ref{eq:1}). The lighter the colour, the higher the coincidence rate is. The colour schemes in the two figures are only relative. One can see that for $s=t$ the two theoretical models give the same pattern.} \label{Fig:comparison}
\end{figure}

For the experiment in figure \ref{Fig:experiment1} both theoretical explanations predict the same interference pattern when looking at two-photon absorption. Also for most other experiments performed so far (see for example \cite{DCS} for double slits, \cite{SEI1,SEI2} for gratings instead of a double slit) either description could be used to explain the obtained results since only same position correlations were measured, and not the detection efficiency. To differentiate the theoretical models an experiment such as the one schematically depicted in figure \ref{Fig:experiment2} can be used.
It is similar to the experiment in figure \ref{Fig:experiment1} in the sense that we ascertain that the state $\ket{\Psi_{{\rm 2,sl}}}$ is generated directly after the double slit. This can be achieved in the same manner as in the experiment previously described. However, instead of looking at same-position, two-photon correlation, we split the beam by an ordinary 50/50 beam splitter (BS) and detect the photons by two detectors that can have arbitrary positions relative the respective optical axis. We only look at the case where both detectors give a click, i.e., we are only looking at coincidences.
We denote $s$ as the position of one of the detectors and $t$ as the position of the other detector. Subsequently the coincidence counts as a function of $s$ and $t$ are recorded. For such an experiment the two theoretical explanations predict different results. The detection probability of the two photons corresponding to (\ref{eq:4}) is \beq P_{{\rm B,sc}}(s,t) \propto \cos^2\left (\frac{k d b s}{ R} \right )\delta_{st}, \label{Eq: Boto photon}  \eeq where $\delta_{st}$ denotes the Kronecker delta. For (\ref{eq:5}), the detection probability becomes \beq P_{{\rm St,sc}}(s,t) \propto \cos^2\left [\frac{k d b}{2 R}( s + t) \right]. \label{Eq: Steuernagel photon}  \eeq In figure \ref{Fig:comparison} the detection coincidence rates are plotted as a function of $s$ (horizontal) and $t$ (vertical). (The lighter the colour, the higher is the coincidence rate.) Figure \ref{Fig:comparison}(a) illustrates the prediction of the model of Boto \etal (\ref{eq:4}) and (\ref{Eq: Boto photon}), where the photons retain spatial correlation. Figure \ref{Fig:comparison}(b) illustrates the case of Steuernagel's description, corresponding to (\ref{eq:5}) and (\ref{Eq: Steuernagel photon}), where the photons can propagate independently after the slit and hence allow an interference pattern even for the case when the two detectors are at different positions. The case where
$s=t$, that is, along the figures' lower left to upper right diagonals, illustrate the results of most the experiments hitherto reported. For this case one can se that both theoretical models predict identical correlation patterns.

\subsection{Performing the experiment}

An experiment similar to that described in figure \ref{eq:2} was recently performed by Peeters \etal \cite{PRE} in another context. The result of
the measurement can be seen in figure 5(a) of their paper \cite{PRE} and the figure is more or less identical to figure \ref{Fig:comparison}(b), indicating that the two photons through the double slit are diffracted independently, but the two-photon character of the state will still be manifested in the coincidence correlation.

Our understanding of the result is that since photons are non-interacting exchange particles, they will diffract and propagate independently in a linear medium. This means that in absence of some additional ``guiding'', the photons prepared in the state $\ket{\Psi_{{\rm 2,sl}}}$ will spread out according to normal diffraction theory, and it will be unlikely to detect the two photons at the same position far from the slit. This is also the case for more complicated states such as the inverse binomial states discussed in \cite{BSS2}. The entanglement will dictate the ensuing $N$ photon interference pattern, but each photon will in principle have its probability amplitude spread over the whole illuminated area, resulting in very low $N$-photon, single detector, correlation. In Appendix B we show that this is also true for the slightly different setup discussed in \cite{Boto}. In other words, Steuernagel seems to be correct in his analysis that both spatial and temporal concentration cannot be achieved at the same time at the required level for quantum lithography \cite{St2}.

In principle, one could try to keep the spatial correlation by making the propagating media non-linear, and, e.g., form spatial solitons \cite{Soliton}.
However, to manage this on a few photon level seems farfetched and not very realistic.

Finally, one may ask if our conclusion is valid also for quantum lithography involving more than two modes \cite{BSS1,BSS2,DSS,KBA}? The answer is yes. Invoking more modes will allow the synthesis of more complicated patterns than a single slit interference pattern superimposed by a sinusoidal modulation, which is what can be achieved by a double slit. By adding more modes one can achieve additional superpositions of modes with different transverse wave vectors. Thus, one can make more involved lithographic patterns. However, if two photons emerging from one of two slits diffract independently, which is what the results of Peeters \etal results indicate, then two (or more) photons will also diffract independently from any one of any number of slits or ``holes'' in a lithographic mask. Therefore, our conclusions about the exposure efficiency are applicable to any lithographic mask when combined with propagation through a linear media.

\section{Can spatial correlation be retained?}
\label{sec:rel}

One might wonder how our paper relates to other papers like \cite{JG} or \cite{PST}, where authors actually predict that $N$-photon absorption proportional to $I$, where $I$ is the normalised classical intensity of single-photon absorption, is theoretically possible? These two papers are also cited by Boto \etal to motivate their argument that quantum lithography works with $I$ instead of an $I^N$ dependence. For the results of these paper to be valid the authors postulate, without giving any specific mechanism, that the photons somehow are transported to exactly the same place, for example \cite{JG} states ``We shall assume that, by focusing or otherwise, the photons of a correlated pair are brought at the same time onto a three-level atom...''. This, however, is not the situation in a typical quantum lithography scheme, since the photons are diffracted independently at a double slit (or more generally, in some lithographic mask) as our paper shows.
In the proposal of Boto \etal \cite{Boto} and the experimental realisation of D'Angelo \etal \cite{DCS}, which does not use a double slit, the photons are spreading independently as an analysis can show (see \ref{app:B}). Therefore, as long as light is propagating through linear media (such as air or vacuum), the results of \cite{JG} and \cite{PST} are not applicable in the existing quantum lithography schemes.

\section{Conclusion}
\label{sec:concl}

We showed that one can experimentally determine the scaling behaviour of the needed exposure time in quantum lithography as a function of the number of photons $N$ and the number of detectors (i.e., pixels) $S$. Such an experiment was recently reported, and it supports the view that photons, spatially entangled at a double slit to subsequently form a sub-wavelength pattern, will diffract independently. The photon state will still form the desired pattern (given that only two-, or in general, $N$-photon coincidences are recorded). However, the experiment unfortunately also indicates that increasing the resolution through the number of entangled photons and by increasing the number of pixels one has to wait exponentially longer time to get the desired interference pattern. The explicit exposure time scaling is $\tau_{{\rm St}}(S,N) \propto S^N$. Therefore, quantum lithography involving a moderate number of pixels unfortunately seems to be impractical even for a rather small number of photons.

\ack
%\section{Acknowledgements}

We acknowledge fruitful discussions with Drs. W.~Peeters and J. S\"{o}derholm. CK acknowledges support from the German National Academic Foundation. The work was supported by the Swedish Research Council (VR) through the Linnaeus Center of Excellence ADOPT, by the Swedish Foundation for International Cooperation in Research and Higher Education (STINT), and by the Knut and Alice Wallenberg Foundation (KAW).

\appendix

\section{Thorough analysis of the theoretical explanations}
\label{app:A}

Only assuming cw-pumping one can write the down-converted two-photon state as
\beq
|\tilde{\Psi}_{\mathrm{crystal}}(\vect{k}^\nichts_1,\vect{k}^\nichts_2) \rangle \propto\tilde{E}_p(\vect{k}^\nichts_1,\vect{k}^\nichts_2)\tilde{\Xi}(\vect{k}^\nichts_1,\vect{k}^\nichts_2) a^\dagger_{\vect{k}^\nichts_1}a^\dagger_{\vect{k}^\nichts_2}\ket{0,0},
\eeq
where the tilde sign $\tilde{}$ denotes Fourier transform and $\vect{k}^\nichts_{1}$ ($\vect{k}^\nichts_{2}$) is the transverse component of the wave vector of the signal (idler) photon. $\tilde{E}$ is the momentum representation of the pump profile in the crystal center plane and $\tilde{\Xi}$ is the phase matching profile. Their exact shape can be specified later. Transfering the above function from momentum to position space gives
\beq
\Psi_{\mathrm{crystal}}(\vect{x}^\nichts_1,\vect{x}^\nichts_2)\propto\int\int \rmd\vect{k}^\nichts_1,\rmd\vect{k}^\nichts_2\tilde{E}_p(\vect{k}^\nichts_1,\vect{k}^\nichts_2)\tilde{\Xi}(\vect{k}^\nichts_1,\vect{k}^\nichts_2)\rme^{-\rmi\vect{k}^\nichts_1\cdot\vect{x}^\nichts_1}\rme^{-\rmi\vect{k}^\nichts_2\cdot\vect{x}^\nichts_2},
\eeq
where we have suppressed the Dirac notation with kets, as the information about subsequent interference lies in the probability amplitudes. The vector $\vect{x}^\nichts_{\ell}=(x^\nichts_{\ell},y^\nichts_{\ell})$, $\ell \in \{1,2\}$ are the transverse coordinates in the crystal center plane and all the integrals here, and in the rest of the paper, are to be taken over the whole real axis.

In the paraxial approximation, i.e., when the angle between signal and idler is not too large, one can write
\beq
\Psi_{\mathrm{slit}}(\vect{x^\eins}_1,\vect{x^\eins}_2)\propto\int\int \rmd\vect{x}^\nichts_1 \rmd\vect{x}^\nichts_2 \Psi_{\mathrm{crystal}}(\vect{x}^\nichts_1,\vect{x}^\nichts_2)h^\prime (\vect{x^\eins}_1,\vect{x}^\nichts_1)h^\prime (\vect{x^\eins}_2,\vect{x}^\nichts_2),
\eeq
where $h^\prime$ is the field propagator \cite{sieg}, which can be specified explicitly later on as well. The propagator $h^\prime$ allows any paraxial system including lenses and other components, but is especially simple for free-space propagation. The transverse coordinates just at the slit plane are denoted $\vect{x^\eins}_{\ell}$. After the double slit we get
\beq
\tilde{\Psi}_{\mathrm{slit}}(\vect{k^\eins}_1,\vect{k^\eins}_2)\propto\int\int \rmd\vect{x^\eins}_1 \rmd\vect{x^\eins}_2 \Psi_{\mathrm{slit}}(\vect{x^\eins}_1,\vect{x^\eins}_2)f(\vect{x}_1^\eins)f(\vect{x}_2^\eins)\rme^{\rmi\vect{k^\eins}_1\cdot\vect{x^\eins}_1}\rme^{\rmi\vect{k^\eins}_2\cdot\vect{x^\eins}_2},
\eeq
where $\vect{k^\eins}_{\ell}$ is the transverse wave vector of the signal ($\ell =1$) or idler ($\ell =2$), just after the slit, but in the slit plane, and where we have assumed that the thickness of the slit is negligible. With $a$ being the width of a slit, $\lambda < a < d$, and $d$ being the distance between the two slits, one can define
\beq
f(\vect{x}_\ell^\eins)=\left\{
\begin{array}{cl}
  1 & \mathrm{for }\quad (d/2)-a/2<|x_{\ell}^\eins|<(d/2)+a/2\\
  0 & \mathrm{else}
\end{array}\right.,
\eeq
i.e., the function is unity at the slit openings and 0 otherwise. (The slit function $f$ of course also depend on the width $a$, but we shall suppress that in the following as we shall be interested in relative, rather than absolute values of fields and intensities.) We have oriented the slits so that they are oriented along the $y_{\ell}^\eins$-direction and have their slit modulation structure in the $x_{\ell}^\eins$-direction.

Finally we transfer the system to position space again and let it propagate to a detection plane so that we get
\beqa
\fl
\nonumber
\Psi_{\mathrm{detect}}(\vect{x}_1^{\drei},\vect{x}_2^{\drei}) & \propto & \int\int\int\int \rmd\vect{x^{\zwei}}_1 \rmd\vect{x^{\zwei}}_2 \rmd\vect{k^{\eins}}_1 \rmd\vect{k^{\eins}}_2 \tilde{\Psi}_{\mathrm{slit}}(\vect{k^\eins}_1,\vect{k^\eins}_2)\\
& &\times h(\vect{x}_1^{\drei},\vect{x^{\zwei}}_1) h(\vect{x}_2^{\drei},\vect{x^{\zwei}}_2) \rme^{-\rmi\vect{k^\eins}_1\cdot\vect{x^{\zwei}}_1} \rme^{-\rmi\vect{k^\eins}_2\cdot\vect{x^{\zwei}}_2},
\eeqa
where $\vect{x^{\zwei}}_{1,2}$ are the transverse coordinates directly after the slit and $\vect{x}_1^{\drei}$ and $\vect{x}_2^{\drei}$ are the transverse coordinates of the detector positions for detecting signal and idler, respectively. Combining all the above formulae, one ends up with
\beqa
\fl
\nonumber
\Psi_{\mathrm{detect}}(\vect{x}_1^{\drei},\vect{x}_2^{\drei}) & \propto & \int \rmd\vect{x^{\zwei}}_1 \int \rmd\vect{x^{\zwei}}_2 \int \rmd\vect{k^{\eins}}_1 \int \rmd\vect{k^{\eins}}_2 \int \rmd\vect{x^{\eins}}_1 \int \rmd\vect{x^{\eins}}_2\\
\nonumber
& &\times \int \rmd\vect{x}^\nichts_1 \int \rmd\vect{x}^\nichts_2 \int \rmd\vect{k}^\nichts_1 \int \rmd\vect{k}^\nichts_2 \tilde{E}_{p}(\vect{k}^\nichts_1,\vect{k}^\nichts_2) \tilde{\Xi}(\vect{k}^\nichts_1,\vect{k}^\nichts_2)\\
\nonumber
& &\times h(\vect{x}_1^{\drei},\vect{x^{\zwei}}_1) h(\vect{x}_2^{\drei},\vect{x^{\zwei}}_2) h^\prime (\vect{x}^\eins_1,\vect{x}^\nichts_1) h^\prime (\vect{x}^\eins_2,\vect{x}^\nichts_2) f(\vect{x}_1^\eins)f(\vect{x}_2^\eins)\\
& &\times \rme^{-\rmi\vect{k^\eins}_1\cdot\vect{x^{\zwei}}_1} \rme^{-\rmi\vect{k^\eins}_2\cdot\vect{x^{\zwei}}_2} \rme^{\rmi\vect{k^\eins}_1\cdot\vect{x^{\eins}}_1} \rme^{\rmi\vect{k^\eins}_2\cdot\vect{x^{\eins}}_2} \rme^{-\rmi\vect{k}^\nichts_1\cdot\vect{x}^\nichts_1} \rme^{-\rmi\vect{k}^\nichts_2\cdot\vect{x}^\nichts_2}.
\eeqa
Without loss of generality we can assume that the detectors are moved only perpendicular to the slits. It is then sufficient to evaluate all the integrals above only in the first $x$ component of the vectors. Thus it suffices to write scalars instead of vectors, so in the paraxial approximation the field propagator can be written as
\beq
h^\prime (x^\eins_1,x^\nichts_1)=\sqrt{-\frac{\rmi}{\lambda \Delta z}} \rme^{\rmi\frac{\pi}{\lambda \Delta z}\left(x^\eins_1-x^\nichts_1\right)^2}
\label{Eq: Field propagator}
\eeq
and correspondingly for the other variables, where $\Delta z=z^\eins-z^\nichts$ is the distance of propagation. If $\Delta z$ is small then $h^\prime (x^\eins_1,x^\nichts_1)$ can be approximated by a delta-function $\delta(x^\eins_1-x^\nichts_1)$.

Now we will make some assumptions. If the slits are very close to the crystal we have $\Delta z\rightarrow0$ and therefore $h^\prime (x^\eins_1,x^\nichts_1)\rightarrow\delta(x^\eins_1-x^\nichts_1)$. Assuming that the slits are narrower than any spatial structure of the two-photon field we can approximate
\beqa
\fl
\nonumber
f(x_1^\eins)f(x_2^\eins) & \propto &  \delta(x_1^\eins-d/2)\delta(x_2^\eins-d/2)+\delta(x_1^\eins-d/2)\delta(x_2^\eins+d/2)\\
& &+ \delta(x_1^\eins+d/2)\delta(x_2^\eins-d/2)+\delta(x_1^\eins+d/2)\delta(x_2^\eins+d/2),
\label{Eq.A9}
\eeqa
where we have assumed that the two slits are identical.

To proceed, the integration over $\vect{k}^\eins_{\ell}$ can be simplified as
\beq
\int \rmd k^\eins_{\ell} \rme^{\rmi k^\eins_{\ell} x^\eins_{\ell}} \rme^{-\rmi k^\eins_{\ell} x^{\zwei}_{\ell}}\rightarrow \delta(x^\eins_{\ell}-x^{\zwei}_{\ell}).
\eeq
Combining all equations and carrying out the integrals leads to
\beqa
\fl
\nonumber
\Psi(x_1^{\drei},x_2^{\drei}) & \propto & \int \rmd k^\nichts_1 \int \rmd k^\nichts_2 \tilde{E}_p(k^\nichts_1,k^\nichts_2)\tilde{\Xi}(k^\nichts_1,k^\nichts_2)\\
\nonumber
& & \times \left[h(x_1^{\drei},d/2)h(x_2^{\drei},d/2) \rme^{-\rmi k^\nichts_1 d/2} \rme^{-\rmi k^\nichts_2 d/2} \right. \\
\nonumber
& & + h(x_1^{\drei},d/2)h(x_2^{\drei},-d/2)\rme^{-\rmi k^\nichts_1 d/2} \rme^{\rmi k^\nichts_2 d/2}\\
\nonumber
& & + h(x_1^{\drei},-d/2)h(x_2^{\drei},d/2)\rme^{\rmi k^\nichts_1 d/2} \rme^{-\rmi k^\nichts_2 d/2} \\
\nonumber
& & + \left. h(x_1^{\drei},-d/2)h(x_2^{\drei},-d/2)\rme^{\rmi k^\nichts_1 d/2} \rme^{\rmi k^\nichts_2 d/2}\right ]\\
\nonumber
& = & c_{11} h(x_1^{\drei},d/2)h(x_2^{\drei},d/2)+c_{12} h(x_1^{\drei},d/2)h(x_2^{\drei},-d/2) \\
& &  + c_{21} h(x_1^{\drei},-d/2)h(x_2^{\drei},d/2)+ c_{22}h(x_2^{\drei},-d/2)h(x_2^{\drei},-d/2) 
\label{Eq.e27}
\eeqa
where, e.g.,
\beqa
\nonumber
c_{11} & = & \int \rmd k^\nichts_1 \int \rmd k^\nichts_2 \tilde{E}_p(k^\nichts_1,k^\nichts_2)\tilde{\Xi}(k^\nichts_1,k^\nichts_2) \rme^{-\rmi k^\nichts_1 d/2} \rme^{-\rmi k^\nichts_2 d/2}\\
& = & \int \rmd u \int \rmd v E_p(u,v)\Xi(d/2-u,d/2-v) .
\label{Eq: Faltning}
\eeqa

In the following, since we discuss interference that require indistinguishability, we shall only consider the preparation of the symmetric state subspace at the slits. This requires symmetry of the functions $E$ and $\Xi$, implying that $c_{11}=c_{22}$ and $c_{12}=c_{21}$. Removing any global phase factor, and ignoring information about the absolute amplitudes, i.e. renormalising $c_{ij}$ to $\tilde{c}_{ij}$, such that $|\tilde{c}_{12}|^2+|\tilde{c}_{11}|^2=1$, we can hence parameterise the amplitudes $\tilde{c}_{11}=\tilde{c}_{22}=\rme(i \varphi) \sin(\alpha/2)$, $\tilde{c}_{12}=\tilde{c}_{21}=\cos(\alpha/2)$. Apart from determining the probability amplitudes, the form of the wave function at the screen is independent of how the pump beam looks like and the particulars of the crystal, if one only imposes symmetry with the respect of the two slits. Of course, the particulars of the pump beam and the crystal determine the absolute magnitudes of $c_{11}$ and $c_{12}$.

We want to stress that the requirement for symmetry ($c_{11}=c_{22}$ and $c_{12}=c_{21}$) is automatically fulfilled if $\tilde{E}$ and $\tilde{\Xi}$ are invariant under the simultaneously exchange of sign in both $k^\nichts_1$ and $k^\nichts_2$. The pump beam profile $\tilde{E}$ is usually a Gaussian beam depending only on the squared sum of the arguments, i.e. $(k^\nichts_1+k^\nichts_2)^2$, which makes it invariant under sign exchange, whereas the phase matching profile $\tilde{\Xi}$ under usual crystal orientation only depends on $(k^\nichts_1-k^\nichts_2)^2$, making it invariant under sign exchange as well. However, it turns out that even under more advanced crystal orientations (the general case is given in equation (2) of \cite{FEV}) the function $\tilde{\Xi}$ stays invariant under sign exchange of their arguments, although the dependence of the argument becomes more complex. Thus our symmetry requirement is almost naturally fulfilled for any crystal systems as long as the slits are symmetrical to the optical axis. 

Above, we assumed that the crystal is sufficiently close to the slit so that $\Delta z \rightarrow 0$. This assumption, however, can be dropped since any $\Delta z$ would transfer $\tilde{\Psi}_{\mathrm{crystal}}(k^\nichts_1,k^\nichts_2)$ to some $\tilde{\Psi}_{\mathrm{slit}}(k^\eins_1,k^\eins_2)$, and from there on, the analysis will be identical, ending up with the expression (\ref{Eq.e27}) in this case too. This expression is in other words very general. The assumptions we cannot drop are that the slits are symmetrically excited by the two-photon field and that the field does not vary appreciably over the width of the slits.

Inserting (\ref{Eq: Field propagator}) into (\ref{Eq.e27}) and considering a NOON state ($\alpha=\pi$) where both photons either take one slit or the other, substituting $2 \pi/\lambda=k$, and assuming a propagation distance $R$ between the slits and the detector, we get
\beq
\Psi_{\mathrm{NOON}}(x_1^{\drei},x_2^{\drei})\propto \rme^{\rmi \frac{k}{2 R} [(x_1^{\drei})^2 + (x_2^{\drei})^2]} \left [ \rme^{-\rmi \frac{k d}{2 R}(x_1^{\drei}+x_2^{\drei})} + \rme^{\rmi \frac{k d}{2 R}(x_1^{\drei}+x_2^{\drei})} \right ].
\eeq
The detection probability at the detectors hence becomes
\beq
P_{\mathrm{NOON}}(x_1^{\drei},x_2^{\drei}) \propto \cos^2\left[\frac{k d}{2 R} (x_1^{\drei}+x_2^{\drei})\right],
\label{Eq: NOON coincidence}
\eeq
or, if $x_1^{\drei}=x_2^{\drei}$, to $P_{\mathrm{NOON}}(x_1^{\drei})\propto \cos^2\left(\frac{k d}{R} x_1^{\drei}\right)$. Equation (\ref{Eq: NOON coincidence}) is of the same form as (\ref{Eq: Steuernagel photon}) (with $b s = x_1^{\drei}$ and $b t = x_2^{\drei}$), but now we arrived at it through a more thorough mathematical analysis.

If we instead consider a state where the photons ``chose'' slit independently, so that $c_{11}=c_{12}=c_{21}=c_{22}$, or $\alpha=\pi/2$ and $\varphi=0$, one obtains the coincidence probability
\beq
P(x_1^{\drei},x_2^{\drei}) \propto \cos^2\left(\frac{k d}{2 R} x_1^{\drei}\right)\cos^2\left(\frac{k d}{2 R} x_2^{\drei}\right),
\label{Eq: Independent slits}
\eeq
or, if $x_1^{\drei}=x_2^{\drei}$, to $P(x_1^{\drei}) \propto \cos^4\left(\frac{k d}{2 R} x_1^{\drei}\right)$. Comparing these expressions shows that the NOON-state leads to a doubling of the period, whereas allowing the photons to ``chose'' slit independently does not.

\begin{figure}
\center
\begin{tabular}{ c c c }
\includegraphics[angle=0, scale=.3]{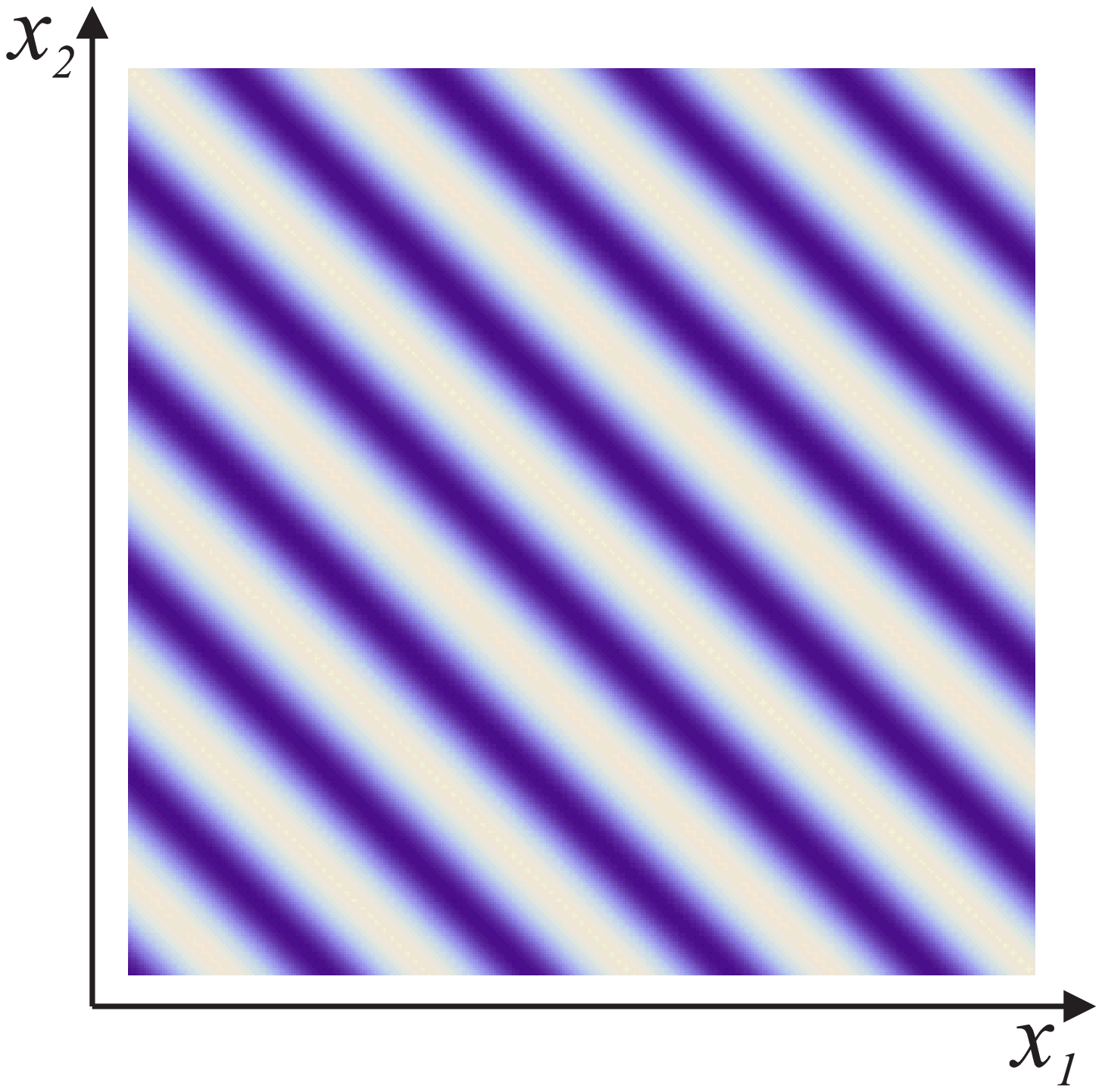} &
\includegraphics[angle=0, scale=.3]{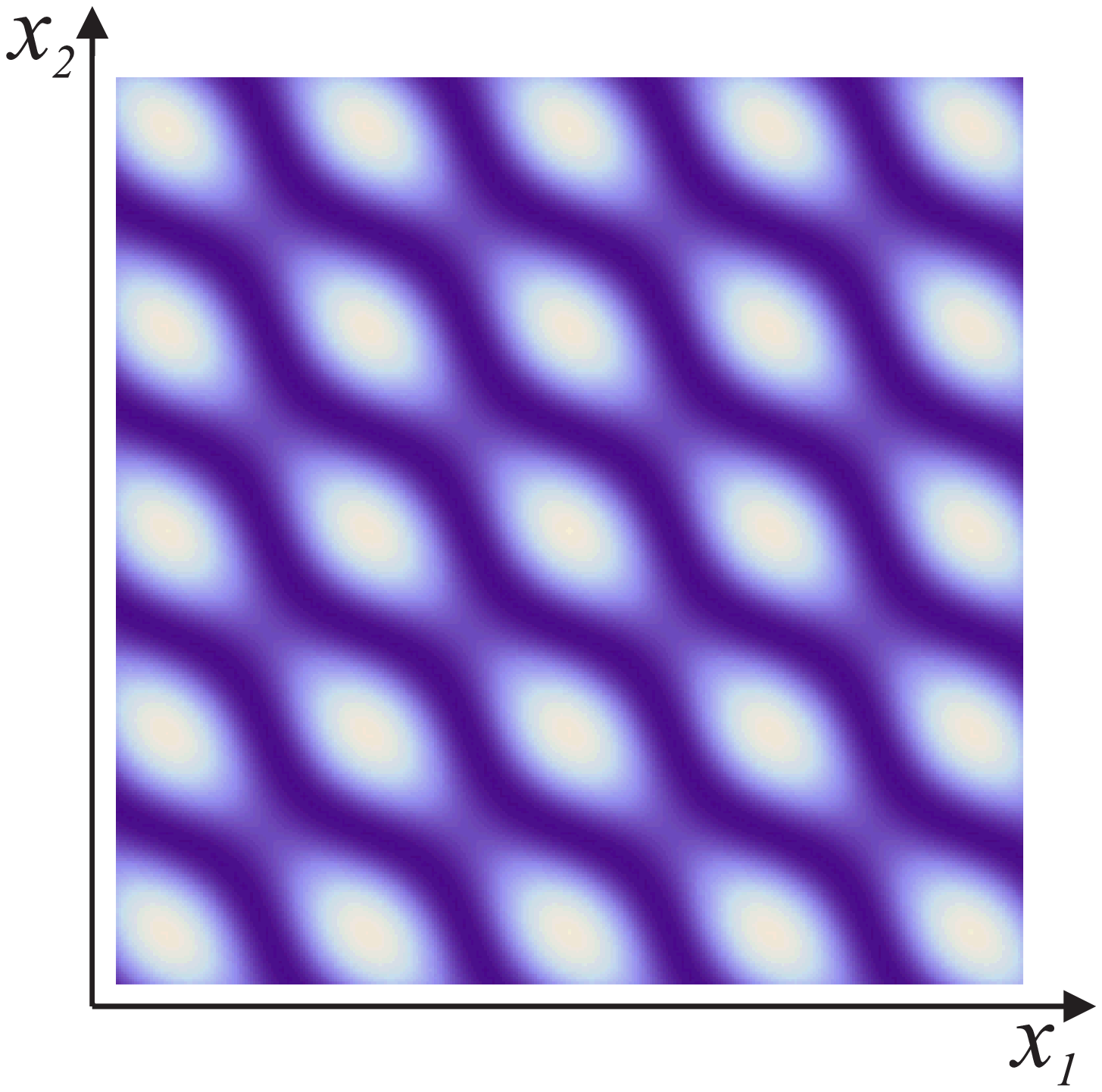} &
\includegraphics[angle=0, scale=.3]{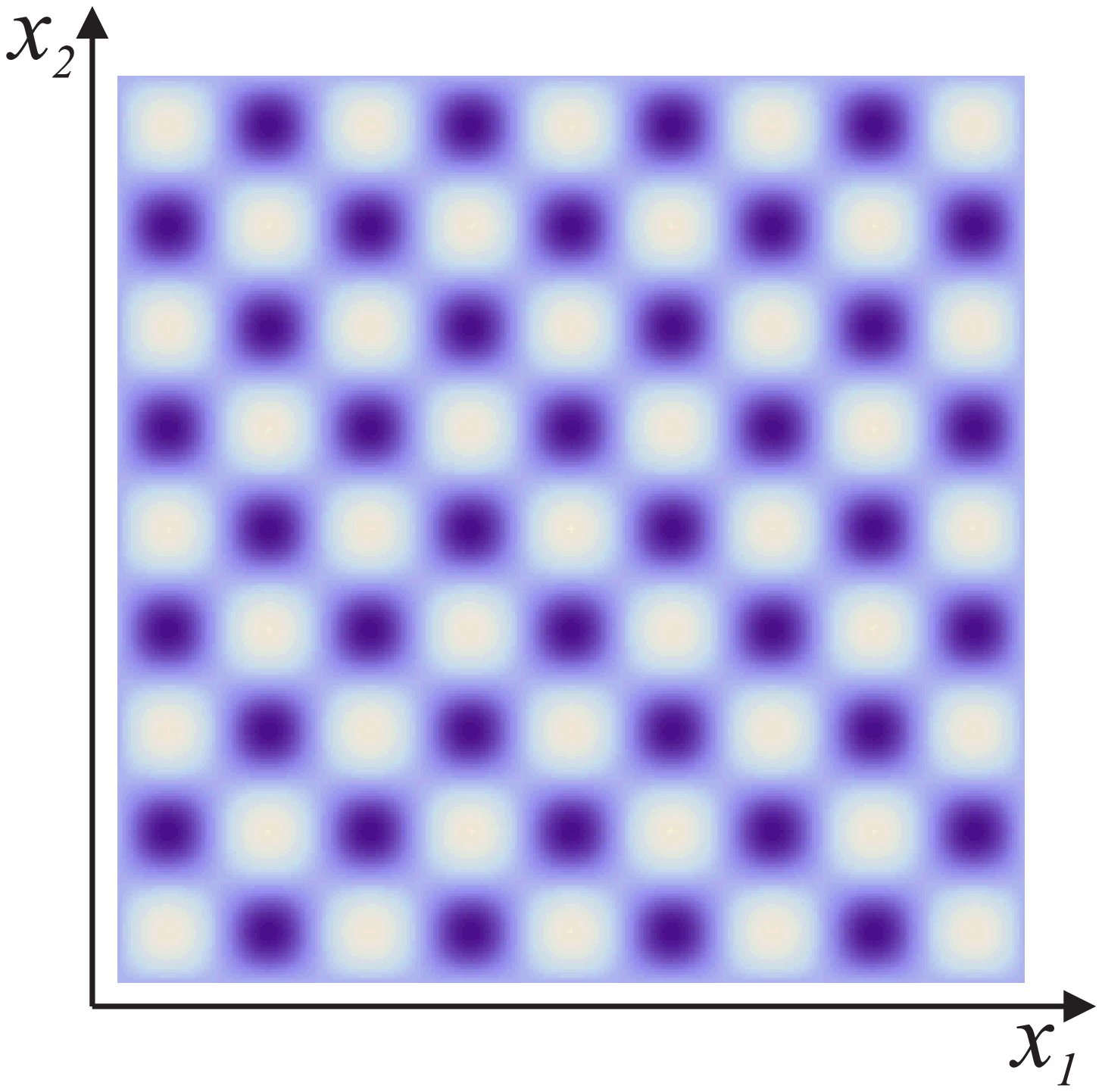}\\
(a) & (b) & (c)\\
\\
\includegraphics[angle=0, scale=.3]{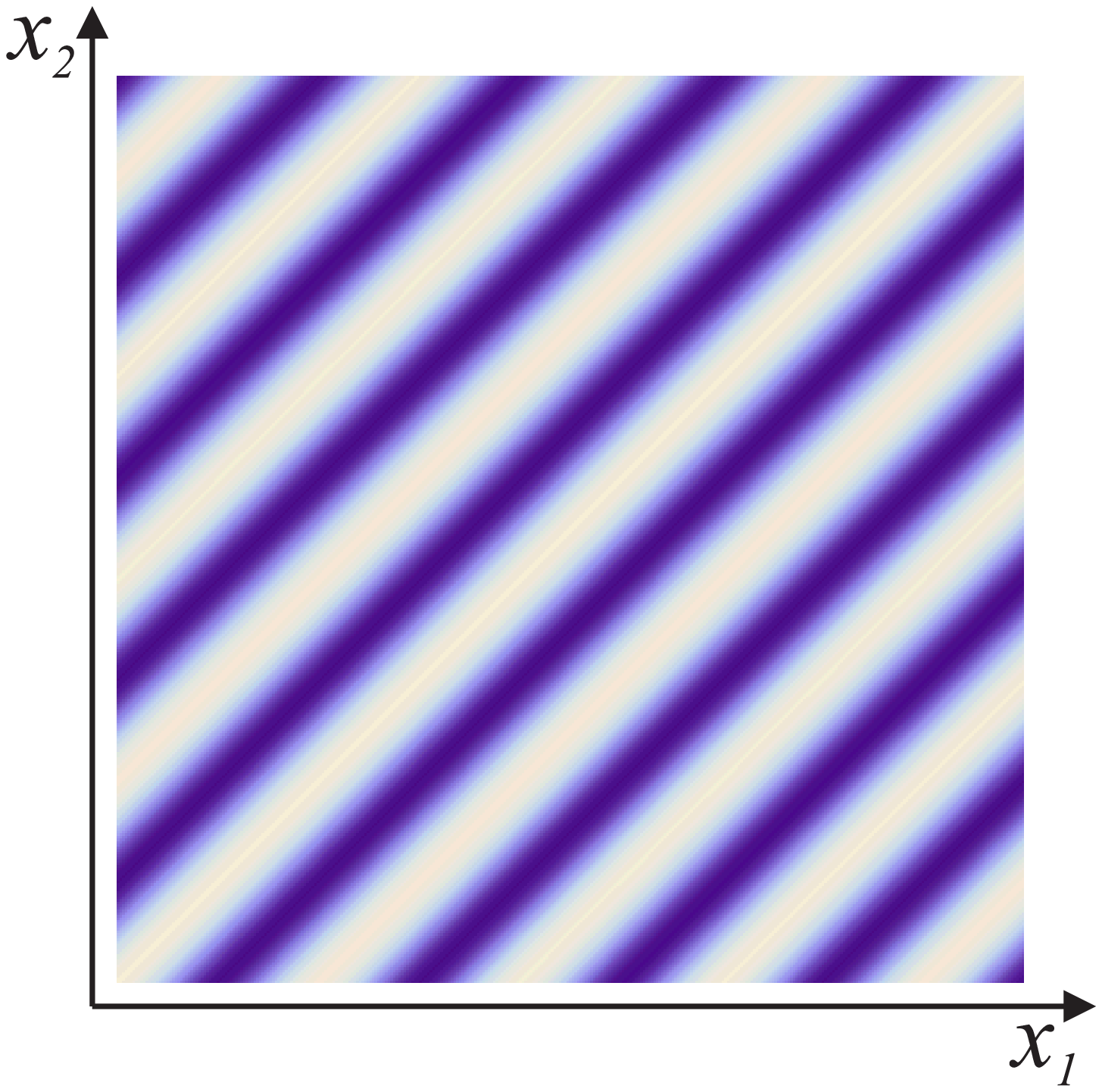} &
\includegraphics[angle=0, scale=.3]{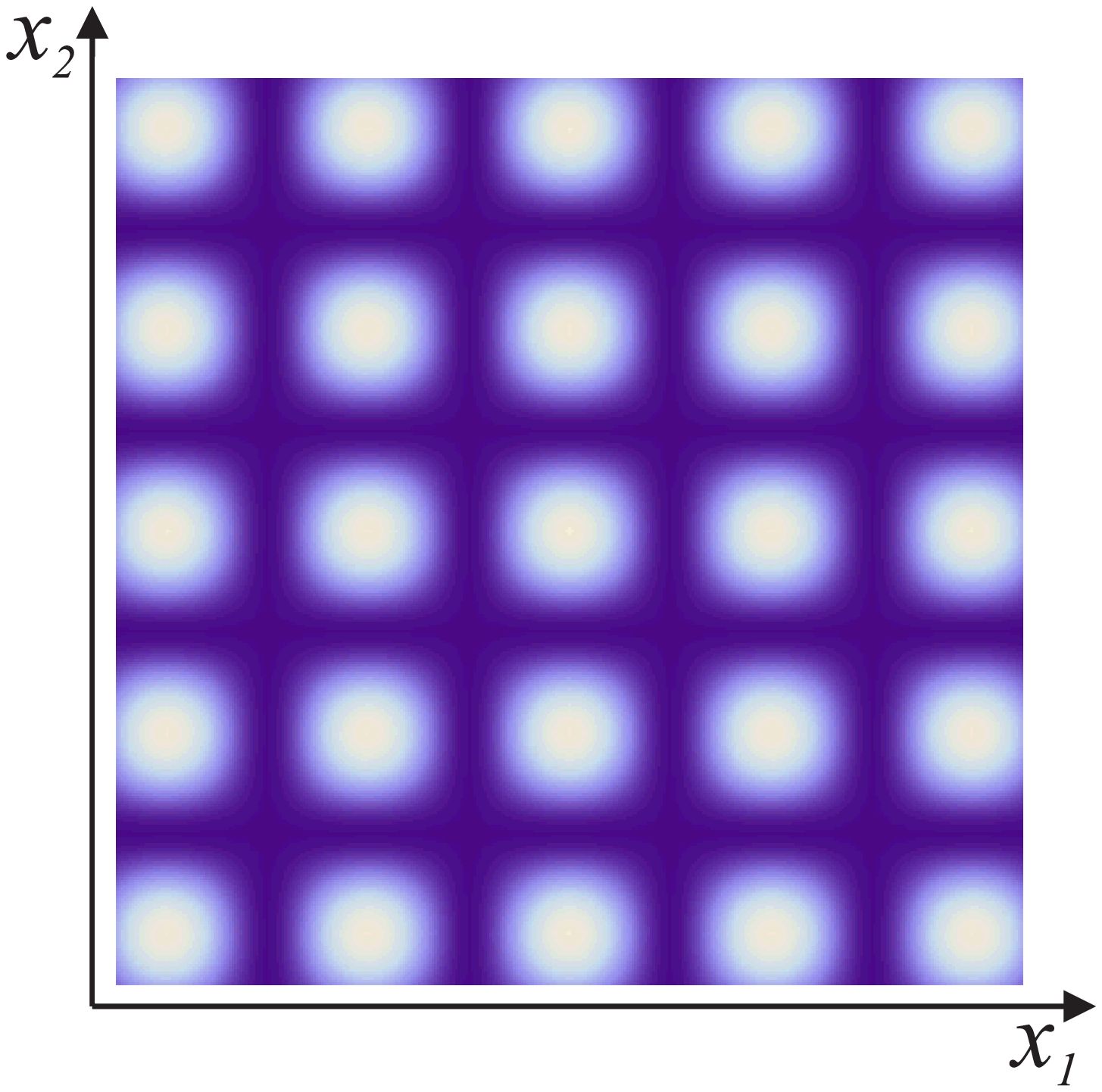} &
\includegraphics[angle=0, scale=.3]{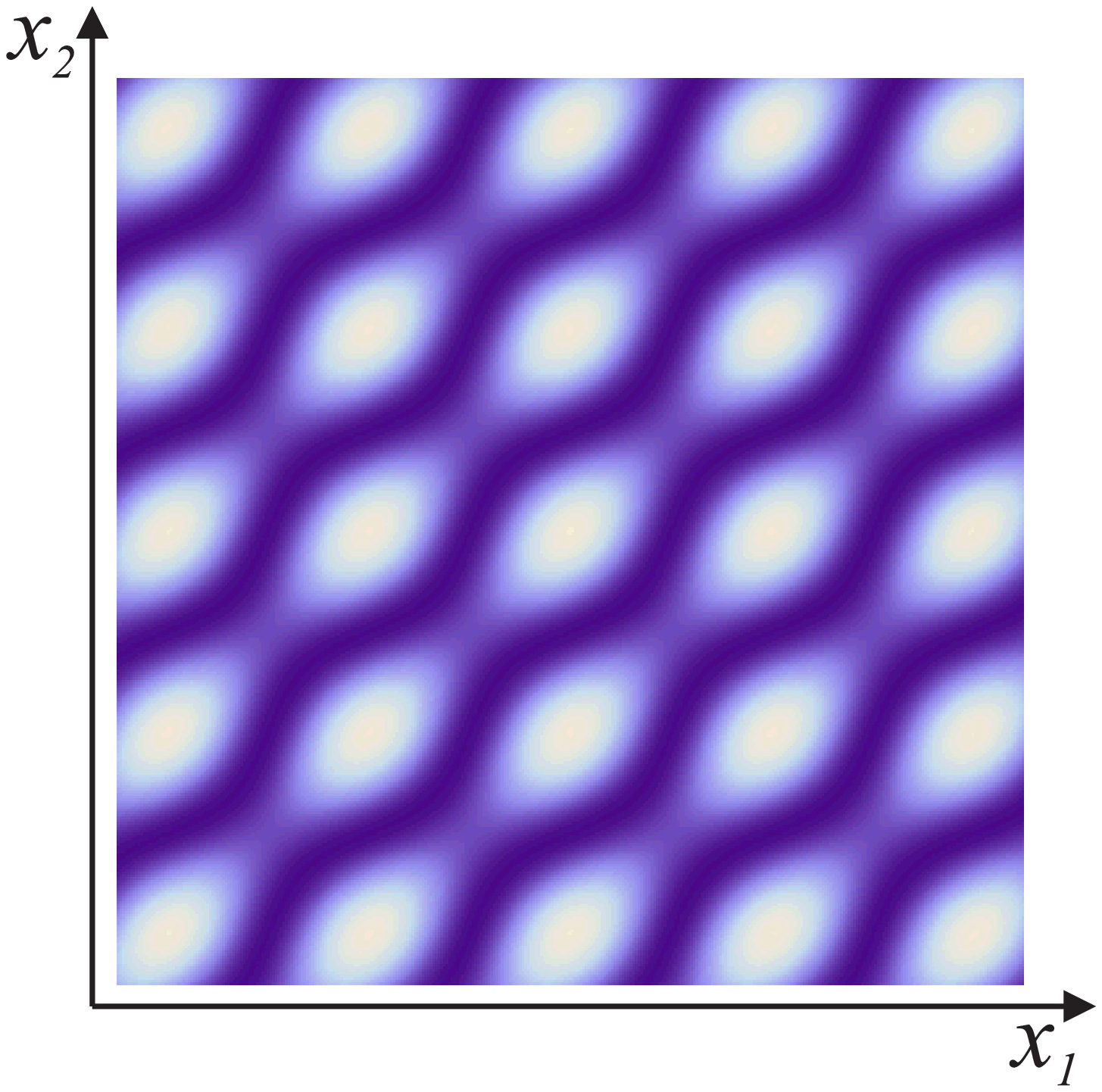}\\
(d) & (e) & (f)\\
\end{tabular}
\caption{
Different plots of (\ref{Eq.e27}) where each axis represents one detector coordinate, i.e., $x_1^{\drei}$ and $x_2^{\drei}$. A lighter colour indicate a higher intensity. The parameters are (a) $\alpha=\pi$, $\varphi=0$, (b) $\alpha=3\pi/2$, $\varphi=0$, (c) $\alpha=\pi/2$, $\varphi=\pi/2$, (d) $\alpha=0$, $\varphi=0$, (e) $\alpha=\pi/2$, $\varphi=0$, and (f) $\alpha=\pi/4$, $\varphi=0$. These calculated plots should be compared to the experimental results of Peeters \etal, presented in figure 5 of their paper \cite{PRE}.}
\label{Fig:comparisontoPeeters}
\end{figure}

Inserting (\ref{Eq: Field propagator}) into (\ref{Eq.e27}) and varying the probability amplitude parameters $\alpha$ and $\varphi$, one obtains figure \ref{Fig:comparisontoPeeters}. These pictures correlate extremely well with figure 5 in the paper of Peeters \etal \cite{PRE} and show that the theory above provides an excellent description of their experiment. (Or rather, that Peeters \etal are to be applauded for their ability to generate the various two-photon, two-slit states they aim for.) 

Note that these figures are not ``ordinary'' interference pictures, as they represent the correlations between two ``point-detectors'' along the same line, and not the intensity as a function of two-dimensional space. The idea of quantum lithography is illustrated in figure \ref{Fig:comparisontoPeeters}(a), that describes NOON-state interference, (\ref{Eq: NOON coincidence}), and should be compared to figure \ref{Fig:comparison}.  Single photon detection as in (\ref{Eq: single photon}) can be seen as if one detector, let's say $x_2$ is kept at a fixed position and only the other detector is moved. In that case one finds four and a half interference maxima along the $x_1$ axis. The two-photons absorption pattern is given by requiring that both detectors have the same coordinate, i.e., $x_1=x_2$. This corresponds to the diagonals in the figures, and going from the lower left to the upper right corner along the diagonal in (a) one counts nine maxima, i.e., a doubling of the number of fringes as compared to the single photon case. As a contrast, one does not see this period doubling for $x_1=x_2$ in (e), which corresponds to (\ref{eq.A.15}), a state without spatial entanglement at the slits. This state will be produced, e.g., if the slits are located far from the crystal. In this case, the state after the slits becomes
\beq
\Psi(x_1^{\drei},x_2^{\drei})=(\ket{2,0} + \sqrt{2}\ket{1,1} + \ket{0,2})/2 = \frac{(a^\dagger + b^\dagger)^2}{2 \sqrt{2}} \ket{0,0},
\label{eq.A.15}
\eeq
where $a^\dagger$ ($b^\dagger$) is the creation operator of a photon through slit 1 (2). Hence, the transmission probabilities are those expected for two independent particles. As can be seen from (\ref{Eq: Independent slits}), the resulting two-photon interference pattern is the product of two single photon interference fringe patterns.

The peculiar pattern in figure \ref{Fig:comparisontoPeeters}(c) is the result of a superposition $(\ket{2,0} - \rmi \sqrt{2}\ket{1,1} + \ket{0,2})/2$ between a NOON state and the state $\ket{1,1}$ in such a way that it cannot be written as the product of two independent photons. Figure \ref{Fig:comparisontoPeeters}(d) describes the interference of the state $\ket{1,1}$ and as expected, it does not show any same position ($x_1^{\drei}=x_2^{\drei}$), two-photon interference but does show different position two-photon correlation fringes. However, what we ask the reader to take home from figure A1 is that Steuernagel's description of quantum lithography describes the experiment of Peeters \etal (correlations are predicted also where $s \neq t$), whereas the theory of Boto \etal does not.

\section{Closer inspection of the original proposal of Boto \etal}
\label{app:B}

We now assume a situation as in the proposal of Boto \etal \cite{Boto} and in the experiment of D'Angelo \etal \cite{DCS}, i.e., two Gaussian beams meeting at a point $z=0$, where the angles between the direction of propagation of the Gaussian beams and the $z$-axis is $\pm \alpha$ and one of the beams is ``above'' the $z$-axis, whereas the other one is ``below'' so that the situation is symmetrical with respect to the $z$-axis. The two positions where the waists of the Gaussian beams are minimal are at a distance $L$ away from the point $x=y=z=0$, where we put the $x$-axis such that the two positions with minimum waists are at $x=\pm L\sin(\alpha)$, $y=0$, and $z=-L\cos(\alpha)$.
The envelope function of a Gaussian beam at this minimum waist position can be written as
\beq
\env_G(\vect{r})=\env_0 \rme^{-\rmi k z}\exp\left(-\rmi k\frac{\vect{r}^2}{2 q}\right),
\eeq
where $\env_0$ is a normalisation constant and
\beq
\frac{1}{q}=\frac{1}{\rho}-\rmi\frac{\lambda}{\pi w^2},
\eeq
where $\rho$ the beam's phase radius of curvature, $w$ the beam waist radius and $\vect{r}$ is the radial distance from the beam's optical axis. Moreover, at the mimimum waist position we have $\rho=\infty$.

The state for $N$ photons in a Gaussian beam is then given by
\beq
|\Psi_0(\vect{r}_1,\vect{r}_2,\ldots,\vect{r}_N) \rangle=\env_G(\vect{r}_1)\env_G(\vect{r}_2)\ldots \env_G(\vect{r}_N) a^\dagger_{\vect{r}_1}a^\dagger_{\vect{r}_2} \ldots a^\dagger_{\vect{r}_N}\ket{0,0, \ldots , 0}.
\label{Eq.e3}
\eeq
Here too, we shall suppress the Dirac notation in the following. The probability amplitude of the state after having propagated a distance $L$ along its axis of propagation is given by
\beqa
\nonumber
\fl
\Psi_L(\vect{r^\einsB}_1,\vect{r^\einsB}_2,\ldots,\vect{r^\einsB}_N) & = & \rme^{-\rmi NkL}\left(\int\right)^N h(\vect{r^\einsB}_1,\vect{r}_1) h(\vect{r^\einsB}_2,\vect{r}_2) \ldots h(\vect{r^\einsB}_N,\vect{r}_N)\\
& & \times \env_G(\vect{r}_1)\env_G(\vect{r}_2)\ldots \env_G(\vect{r}_N) d\vect{r_1}d\vect{r_2}\ldots d\vect{r_N},
\label{Eq.e4}
\eeqa
where $h(\vect{r}^\einsB,\vect{r})$ is the field propagator. For Gaussian beams propagating a distance $L$ these integrals can be solved and one can write
\beq
\Psi_L(\vect{r^\einsB}_1,\vect{r^\einsB}_2,\ldots,\vect{r^\einsB}_N)=\rme^{-\rmi NkL} \env^\einsB_G(\vect{r^\einsB}_1)\env^\einsB_G(\vect{r^\einsB}_2)\ldots \env^\einsB_G(\vect{r^\einsB}_N),
\label{Eq.e5}
\eeq
where \cite{sieg}
\beq
\env^\einsB_G(\vect{r^\einsB})=\env_0 \sqrt{\frac{1}{1+L/q}}\exp\left(-\rmi k\frac{\vect{r^\einsB}^2}{2 (q+L)}\right).
\label{Eq.e6}
\eeq

Coming back to the case of the proposal and the experiment, the detector is placed at $y=z=0$ with $x$ as the free variable and not with $\vect{r^\einsB}$ as in our description above. We therefore have to change the coordinates in (\ref{Eq.e6}). To do that, however, only photons arriving at $x=0$ will have traveled the distance $L$, whereas photons arriving at any other position will have travelled the distance $L\mp\Delta L(x)$, where the minus-sign is for the case of the Gaussian beam coming from above the $z$-axis and the plus-sign for the case, where the Gaussian beam comes from below the $z$-axis. One has the following geometrical relations:
\beq
\sin(\alpha)=\Delta L/r^\einsB,\; \cos(\alpha)=x/r^\einsB,\; \tan(\alpha)=\Delta L/x,
\label{Eq.e7}
\eeq
where $r^\einsB$ is the coordinate of $\vect{r^\einsB}$ which is in the plane spanned by $x$ and $z$ (the other coordinate of $\vect{r^\einsB}$ is parallel to $y$ and will not change by our coordinate transformation). From now on we will for simplicity, but without loss of generality, set $y=0$, since the detector is placed at $y=0$ and the coordinate systems of both two Gaussian beams have their $y$-axis oriented in the same direction. Equation (\ref{Eq.e6}) then transforms to
\beq
\env^\einsB_G(x)=\env_0 \sqrt{\frac{1}{1+[L\mp x\sin(\alpha)]/q}}\exp\left(-\rmi k\frac{[x\cos(\alpha)]^2}{2 [q+L\mp x\sin(\alpha)]}\right).
\label{Eq.e8}
\eeq

Since $q$ in general is a complex number both the square root and the exponential function in (\ref{Eq.e8}) are complex numbers. To be able to calculate any interference effects between the two Gaussian beams we will split the two functions explicitly into their real and imaginary parts. Introducing the dimensionless parameter $\beta=\lambda^2 L^2/(\pi^2 w^4)$ and looking only at the case when the minimum waist of the Gaussian beam is not too close to the detectors, i.e., $L\gg x\sin(\alpha)$, we can simplify the above expressions to
\beq
\env^\einsB_G(x)=\env_0 (1+\beta)^{-1/4} \rme^{\frac{\rmi}{2}\arctan(\sqrt{\beta})} \rme^{-\frac{k \sqrt{\beta} x^2\cos^2(\alpha) }{2 L(1+\beta^{-1})}} \rme^{-\rmi\frac{k x^2\cos^2(\alpha)}{2L(1+\beta^{-1})}} \rme^{\mp \rmi\frac{k x^3 \cos^2(\alpha)\sin(\alpha)}{2L^2(1+\beta^{-1})}}.
\eeq
The second and the last exponential-functions on the right hand side are the important terms here. The second term will be the Gaussian envelope and the last term will depend on whether a photon originated from the upper or the lower path. Having changed the coordinates of $\env^\einsB_G$ we have to be careful to also change the expression $\rme^{(-i N k L)}$ in (\ref{Eq.e4}) and (\ref{Eq.e5}), since $L$ there differs as well, depending on where the detector is placed. This leads to the replacement
\beq
\rme^{-\rmi N k L}\rightarrow \rme^{-\rmi k \sum\limits_{i=1}^{N} L_i (x)}=\rme^{-\rmi N k L} \rme^{-\rmi k \sin(\alpha)\sum\limits_{i=1}^{N}\mp x_i},
\label{Eq.e12}
\eeq
where the $\mp$-sign is defined in the same manner as above.

All the calculations so far have considered a single Gaussian beam. In the proposal of Boto \etal, and in the experiment by D'Angelo \etal, we have a NOON-state, i.e., a superposition of $N$ photons in an upper Gaussian beam and zero in a lower and vice versa. For such a case, the probability amplitude $\Psi$ at the place of the detectors ($z=0$) considering the results above can, after some algebra, be written as
\beqa
\fl
\nonumber
\Psi_{\mathrm{NOON}}(x_1,x_2,\ldots x_N) & = & \sqrt{2}\left[\frac{\env_0}{(1+\beta)^{1/4}}\right]^N \rme^{-\rmi N k L} \rme^{\frac{\rmi}{2}N\arctan(\sqrt{\beta})} \rme^{-\frac{k \cos^2(\alpha) \sqrt{\beta}\sum\limits_{i=1}^{N}x_{i}^2}{2 L(1+\beta^{-1})}} \rme^{-\rmi\frac{k \cos^2(\alpha)\sum\limits_{i=1}^N x_i^2}{2L(1+\beta^{-1})}}\\
& & \times \cos\left(k \sin(\alpha)\sum\limits_{i=1}^{N}x_i-\frac{k \cos^2(\alpha)\sin(\alpha)\sum\limits_{i=1}^N x_i^3}{2L^2(1+\beta^{-1})}\right).
\label{Eq.e13}
\eeqa
If we only look at the case of detecting all the $N$ photons at the same place (i.e., $x_1=x_2=\ldots=x_n=x$) we get the detection probability
\beqa
\fl
\nonumber
P_{\mathrm{NOON}}(x,x,\ldots x) & = & 2\left[\frac{\env_0}{(1+\beta)^{1/4}}\right]^{2N} \rme^{-\frac{k \cos^2(\alpha) \sqrt{\beta}N x^2}{L(1+\beta^{-1})}}\\
& & \times\cos^2\left[k \sin(\alpha)N x-\frac{k \cos^2(\alpha)\sin(\alpha) N x^3}{2L^2(1+\beta^{-1})}\right].
\label{Eq.e14}
\eeqa
To see any interference effects we have to demand that the cosine-function oscillates faster than the exponential envelope function vanishes. This is fulfilled either for
\beq
k \sin(\alpha)N x>\frac{k \cos^2(\alpha) \sqrt{\beta}N x^2}{L(1+\beta^{-1})}
\eeq
or for
\beq
\frac{k \cos^2(\alpha)\sin(\alpha) N x^3}{2L^2(1+\beta^{-1})}>\frac{k \cos^2(\alpha) \sqrt{\beta}N x^2}{L(1+\beta^{-1})}.
\eeq
The above relations are fulfilled if we have $x<\frac{(1+\beta^{-1})\sin(\alpha)}{\cos^2(\alpha)}\frac{\pi w^2}{\lambda}$ or $x>2\pi w^2/[\lambda\sin(\alpha)]$, where we have used the definition of $\beta$. For $\alpha>\pi/4$ one of the two conditions is always fulfilled. To get a feel for these terms, we can rewrite the cosine function in (\ref{Eq.e14}) as
\beq
\cos^2\left[2\pi\sin(\alpha) N \frac{x}{\lambda}-\frac{2\pi\cos^2(\alpha)\sin(\alpha)N}{2(1+\beta^{-1})L^2/\lambda^2}\left(\frac{x}{\lambda}\right)^3\right]
\eeq
and look at a typical example where we set $\alpha=30^\circ, \lambda=1\;\mu\mathrm{m}, L=10\; \mathrm{cm}, w=1\;\mathrm{mm}$ and $x=0.1\;\mathrm{mm}$. For these parameters the pre-factor of a $x/\lambda$-term in the cosine function argument is 14 orders of magnitude higher than the pre-factor of the $(x/\lambda)^3$ term, so that the latter can safely be neglected in this case.

Irrespective of which condition is fulfilled, the distance between two interference fringes scales with $1/N$ when going from 1 to $N$ photons and the width of the envelope function scales with $1/\sqrt{N}$. This is the behaviour predicted by Boto \etal. However, if we look again at (\ref{Eq.e13}) for the case $N=2$ and with the assumption that $L$ is sufficiently large, so that the $x^3$-term in the cosine function can be neglected, we get
\beq
P(x_1,x_2)=2\frac{\env_0^4}{1+\beta} \rme^{-\frac{k \cos^2(\alpha) \sqrt{\beta} (x_1^2+x_2^2)}{L(1+\beta^{-1})}}\cos^2\left[k \sin(\alpha) (x_1+x_2)\right].
\label{Eq.e17}
\eeq
This function produces the same kind of pattern as in Steuernagel's model and shows that our somewhat simple description, (\ref{eq:5}) not only describes the case of the double-slit experiment, but describes also the case of a NOON-state shared between two Gaussian beams, such as the setup-up of D'Angelo \etal. In the derivation of (\ref{Eq.e17}) we have assumed nothing but Gaussian beams and that the original state is in a NOON-state distributed over the two arms.

We will now redo the previous analysis, but now assuming that all the photons are always produced at the same spatial coordinate. Instead of (\ref{Eq.e3}) we use
\beq
\fl
\Psi_0\left(\vect{r}_1,\vect{r}_2,\ldots,\vect{r}_N\right)=\env_0^\prime \rme^{-\rmi k z N}\rme\left(-\rmi k\frac{\vect{r}_1^2}{2q}\right) \delta(\vect{r}_1-\vect{r}_2)\delta(\vect{r}_1-\vect{r}_3)\ldots\delta(\vect{r}_1-\vect{r}_N)
\label{Eq.B16}
\eeq
as our input state probability amplitude, where $\env_0^\prime$ is the new normalisation factor. This state is at the origin of the theory of Boto \etal. The implication is that the first photon can essentially be found anywhere within the beam's Gaussian envelope, but the rest of the photons will subsequently be found at the same position as the first.

 In the same manner as before we can calculate the state after a distance $L$ with help of the propagators. This leads to
\beqa
\fl
\nonumber
\Psi_L(\vect{r^\einsB}_1,\vect{r^\einsB}_2,\ldots,\vect{r^\einsB}_N) & = & \env_0^\prime  \rme^{-\rmi N k L} \sqrt{\frac{1}{N+L/q}} \sqrt{\frac{\rmi}{\lambda L}}^{N-1}\\
& & \times \exp\left(-\rmi\frac{k}{2L}\left[\frac{\left(\sum\limits_{i=1}^{N}\vect{r}_i^\einsB\right)^2}{N+L/q}-\sum\limits_{i=1}^{N}\vect{r}_i^{\einsB 2}\right]\right),
\label{Eq.B17}
\eeqa
which, by looking at the case where all the photons are detected at the same place, i.e., $\vect{r}_1^\einsB=\vect{r}_2^\einsB=\ldots=\vect{r}_N^\einsB:=(x,0)$ and assuming the NOON-state leads to
\beqa
\nonumber
\fl
\Psi_{\mathrm{NOON}}(x) & = & \sqrt{2} \env_0^\prime \rme^{-\rmi N k L} \left(\frac{1}{N^2+\beta}\right)^{1/4} \sqrt{\frac{\rmi}{\lambda L}}^{N-1} \rme^{\frac{\rmi}{2}\arctan\left(\frac{\sqrt{\beta}}{N}\right)} \rme^{\rmi\frac{k N^2 \cos^2(\alpha) x^2 (N-1)}{2L(\beta+N^2)}}\\
& & \times \rme^{-\frac{k N^2 \sqrt{\beta} x^2 \cos^2(\alpha)}{2L(\beta+N^2)}}\cos\left[k\left(\sin\alpha\right) N x\right],
\eeqa
where we did the same replacements as in (\ref{Eq.e7}) and (\ref{Eq.e12}), setting $y=0$ as before, and ignoring terms with higher orders than $x^2$ as before.

This gives the $N$-photon detection probability
\beq
P_{\mathrm{NOON}}(x) = 2 \env_0^{\prime 2}\sqrt{\frac{1}{N^2+\beta}} \left(\frac{1}{\lambda L}\right)^{N-1} \rme^{-\frac{k N^2 \sqrt{\beta} x^2 \cos^2(\alpha)}{L(\beta+N^2)}}\cos^2\left[k\left(\sin\alpha\right) N x\right].
\label{Eq.eII.5}
\eeq
This expression should be compared with (\ref{Eq.e14}), where the $x^3$-term is neglected. One sees immediately that both equations lead to the same distance between nearby peaks, given by the cosine-term. (This distance halving is what has been mainly been looked for in experiments so far.) Writing the equivalent to (\ref{Eq.e17}) when starting with (\ref{Eq.B16}) and (\ref{Eq.B17}) leads to
\beq
P(x_1,x_2)=2 \env_0^{\prime 2}\sqrt{\frac{1}{4+\beta}} \frac{1}{\lambda L} e^{-\frac{k\cos^2(\alpha)\sqrt{\beta}\left(x_1+x_2\right)^2}{L(\beta+4)}}\cos^2\left[k\sin(\alpha)(x_1+x_2)\right].
\label{Eq.B20}
\eeq
Thus, in neither (\ref{Eq.e17}) nor (\ref{Eq.B20}) the photons are constrained to arrive at the same place as assumed by Boto \etal.

A difference, between (\ref{Eq.e14}) and  (\ref{Eq.eII.5}) is the exponential function. If one chooses $L$ and $w$ carefully so that one has $\beta\gg 1$, the argument of the exponential-function is proportional to $\beta^{1/2}\propto w^{-2}$ in (\ref{Eq.e14}), but proportional to $\beta^{-1/2}\propto w^2$ in  (\ref{Eq.eII.5}). Since the argument of the exponential-function determines the number of fringes one can see, one could, by changing $w$, determine which of the states one has produced, since in one case one would see more fringes, but in the other one less. This may be of interest for further experimental work in this field.

\end{document}